\documentclass[fleqn]{2023SCGE}
\usepackage[normalem]{ulem}
\usepackage[retainorgcmds]{IEEEtrantools}

\newcommand{\ud}{\mathrm{d}}

\usepackage{cancel}

\begin{document}

\ensubject{subject}

\ArticleType{Article}
\SpecialTopic{SPECIAL TOPIC: New Challenges and New Physics in Future Cosmology}
\Year{2025}
\Month{August}
\Vol{68}
\No{8}
\DOI{10.1007/s11433-025-2700-y}
\ArtNo{280409}
\ReceiveDate{January 18, 2025}
\AcceptDate{May 26, 2025}
\OnlineDate{June 25, 2025}

\title{
Hydrodynamics of ultralight complex scalar field dark matter
and its impact on the growth of structure
}
{Hydrodynamics of ultralight complex scalar field dark matter
and its impact on the growth of structure
}

\author[1]{Qi Yang}{}%
\author[1]{Bohua Li}{{bohuali@gxu.edu.cn}}
\author[2]{Paul R. Shapiro}{}

\AuthorMark{Q. Yang}

\AuthorCitation{Qi Yang, Bohua Li, and Paul R. Shapiro}

\address[1]{Guangxi Key Laboratory for Relativistic Astrophysics, 
School of Physical Science and Technology,  \\
Guangxi University, Nanning, 530004, China}
\address[2]{Department of Astronomy, The University of Texas at Austin, Austin, 
TX 78712, USA}


\abstract{
The mass window of ultralight axion dark matter
motivated by suppressing the growth of structure on subgalactic scales,
$m\sim 10^{-22}\,\mathrm{eV}$, is now severely constrained
by various observation data (e.g. Lyman-$\alpha$ forest). 
As an attempt to reopen this mass window,
we investigate an alternative ultralight dark matter candidate,
the complex scalar field dark matter (SFDM).
We derive the relativistic hydrodynamics of the complex SFDM
in the framework of cosmological perturbation theory.
Our formalism contains two novel ingredients
uniquely associated with the complex SFDM model:
the Eckart frame defined by the conserved Noether current,
and the stiff gauge condition, $c_s^2\equiv (\delta P/\delta\rho)|_s=1$.
In the Eckart frame, the complex SFDM is effectively
an imperfect fluid with a dissipative energy flux, 
distinguishing itself from axion dark matter.
The energy flux can affect the growth of density fluctuations dynamically.
Meanwhile, we apply the stiff gauge condition to find new constitutive equations
for the complex SFDM.
We revisit the homogeneous evolution of the complex SFDM
and present illustrative early-stage solutions
for perturbations of the complex SFDM in a simplified setting.
We demonstrate the effects of varying the model parameters
on the evolution of the perturbation variables.
}



\maketitle


\begin{multicols}{2}

\section{Introduction}\label{sec:intro}

Ultralight dark matter has been one of the most studied dark matter candidates
over the past few decades 
\cite{1990PhRvL..64.1084P, 2000PhRvL..85.1158H, 
2006JHEP...06..051S, 2010PhRvD..81l3530A, 2010PhRvD..82j3528M,
2014NatPh..10..496S, 2014PhRvL.113z1302S, 2014PhRvD..89h3536L,2015PhRvD..91j3512H, 2017PhRvD..95d3541H};
see \cite{2016PhR...643....1M,2021ARA&A..59..247H,2021A&ARv..29....7F}
for recent reviews.
Comprised of light bosons ($m \ll 1\,\mathrm{eV}$),
cosmological ultralight dark matter has large occupation numbers in its phase space
and hence can be described by a coherent classical scalar field.
Ultralight scalar field dark matter (SFDM) of masses around
$\sim 10^{-22}\,\mathrm{eV}$ is particularly interesting
since it can suppress the growth of structure on sub-kiloparsec scales
and produce a flat solitonic core at the centers of dark matter halos,
due to its wave nature (e.g., \cite{2000PhRvL..85.1158H, 2014PhRvL.113z1302S}).
This may help resolve the small-scale controversies
of the cold dark matter (CDM) model
\cite{2014Natur.506..171P, 2015PNAS..11212249W,2017ARA&A..55..343B}.
While the validity of these controversies is still under debate
\cite{2015MNRAS.452.3650O, PhysRevLett.119.111102, 2023MNRAS.521.1316R}, 
it is important to understand the dynamical implications
of relevant alternative dark matter candidates like ultralight SFDM.
In the meantime, the mass window of $m\sim 10^{-22}\,\mathrm{eV}$ for ultralight SFDM
is also preferred by the string axiverse scenario
combined with current cosmological data \cite{2019PhRvD..99f3517V}.

However, recent Lyman-$\alpha$ forest observations
have placed strong bounds on the mass of ultralight SFDM, 
$m > 2\times 10^{-20}\,\mathrm{eV}$ (95\% credible limit)
\cite{2017PhRvL.119c1302I, 2021PhRvL.126g1302R},
thus threatening to close the mass window of $\sim 10^{-22}-10^{-19}\,\mathrm{eV}$
and undermine the motivation for ultralight SFDM to resolve the CDM problems.
In addition, future cosmological data
such as 21-cm brightness temperature fluctuations
may further constrain the mass window for SFDM
\cite{2021ApJ...913....7J,2022PhRvD.106f3504F}.
On subgalactic scales, the subhalo mass function
measured by strong lensing and stellar streams
also places competitive constraints on the boson mass,
$m \gtrsim 10^{-21}\,\mathrm{eV}$ \cite{2020PhRvD.101l3026S,2020PhRvD.101j3023B}. 
Thus, in view of existing and future observations, 
ultralight SFDM of masses $\sim 10^{-22}\,\mathrm{eV}$ is now severely challenged
as a potential remedy for the small-scale CDM problems. 

In order to save this appealing mass window for ultralight SFDM, 
we remark that the above constraints are based on
the simplest real scalar field model with no self-interactions
(dubbed ``fuzzy dark matter'', FDM), whose particle prototype
is provided by ultralight string axions
\cite{2006JHEP...06..051S,2010PhRvD..81l3530A}.
Therefore, observational constraints like Lyman-$\alpha$ forest
need to be reexamined if one considers variants of the simplest FDM model
with different predictions for the growth of structure.
It is possible that such variants with masses in the desired mass window
may be able to fit the observation data.

In this work, we consider the complex scalar field dark matter model
\cite{1990PhRvL..64.1084P, 2002PhLB..545...17B, 2011PhLB..696..315B, 2014PhRvD..89h3536L,2017PhRvD..96f3505L}
and assume that the complex SFDM makes up the total dark matter sector. 
Complex scalar fields exist ubiquitously in particle physics,
e.g., the Higgs mechanism, 
the Peccei-Quinn symmetry breaking \cite{1977PhRvL.38.P.Q.}, etc. 
In astrophysics, complex scalar fields have been applied to
the formation of primary black holes, boson stars, and cosmological dark matter
\cite{1985MNRAS.215..575K,1990PhRvD..42..384S, 1994PhRvL..72.2516S,1995PhRvD..51.5698S,1997PhRvD..55.7440J,
2002PhRvD..65h3514A,2006PhRvD..74d3516A, 2022PhRvD.105l3534F}.

The purpose of this paper is to lay the theoretical foundation
for studying the growth of cosmic structure with the complex SFDM model,
in the framework of cosmological perturbation theory.
The global U(1) symmetry of a complex scalar field
gives rise to a conserved Noether charge, 
equal to the difference between the number of particles and antiparticles. 
This Noether charge is absent in the FDM or any real SFDM model, 
thus a distinct signature of the complex SFDM.
Furthermore, the comoving frame of the conserved Noether current
provides a new approach to the relativistic hydrodynamics of ultralight SFDM.
In this frame (called the Eckart frame \cite{1940PhRv...58..919E}), 
we show that the complex SFDM is equivalent to an imperfect fluid
that contains an energy flux term, 
conceptually similar to the conduction term in heat transfer.
The energy flux induces dynamical effects on the evolution of density perturbations.
Finally, we identify a novel gauge condition motivated by the complex SFDM,
the \emph{stiff} gauge, in which the SFDM sound speed equals the speed of light.
We derive the constitutive equations for the complex SFDM
with the aid of the stiff gauge.
We present illustrative numerical solutions to the perturbation equations.

The paper is organized as follows. 
In \cref{section2}, we describe the hydrodynamic formalism
of the complex SFDM model, introducing new signatures
distinct from the real, axion-like model.
In \cref{section4}, we present exact numerical results
for the background evolution of the complex SFDM. 
In \cref{section5}, we discuss the cosmological perturbation equations
for the complex SFDM
and present approximate early-stage solutions in the synchronous gauge.
We conclude in \cref{sec:conclusion}.

\section{Formalism: complex scalar field}\label{section2}

We begin with the following action 
that describes a massive complex scalar field minimally coupled to gravity,
\begin{equation}\label{eq.2.1-1}
  S = \int \ud^4x \sqrt{-g}\,(g^{\mu\nu}\partial_\mu\phi^* \partial_\nu \phi - m^2\phi^*\phi),
\end{equation}
where we adopt the metric signature $(+,-,-,-)$
and natural units, $c = \hbar = 1$, throughout the paper.
The complex SFDM obeys the Klein-Gordon (KG) equation,
\begin{equation}\label{eq.2.1-2}
   \frac{1}{\sqrt{-g}} \frac{\partial}{\partial x^\mu} 
   (\sqrt{-g} \, g^{\mu\nu}\frac{\partial \phi}{\partial x^\nu}) + m^2\,\phi = 0,
\end{equation}
as its equation of motion (EoM).

In linear cosmological perturbation theory, 
the spacetime geometry is described by 
the Friedmann-Lemaitre-Robertson-Walker (FLRW) metric,
\begin{equation}\label{eq.2.1-3}
\begin{split}
  \ud s^2 =  & (1+2\Psi)\,\ud t^2 + 2a\,\partial_iB\,\ud x^i \ud t \\
    & - a^2 \left[(1-2\Phi)\,\delta_{ij} - 2\left(\partial_i\partial_j-\frac{1}{3}\delta_{ij}\,\nabla^2\right)E\right] \ud x^i \ud x^j,
\end{split}
\end{equation}
where $a(t)$ is the scalar factor and $\Psi$, $B$, $\Phi$, $E$ 
are all scalar functions of $(t,\vec x)$ defined in 
const-$t$ hypersurfaces.
In this paper, we are only concerned with scalar-mode perturbations
relevant to the density fluctuations and the formation of large-scale structure.
Meanwhile, we can split the complex scalar field $\phi$ 
into a homogeneous part and a perturbed part, 
$\phi(t,\vec x) = \bar{\phi}(t) + \delta\phi(t,\vec x)$. 
Therefore, the KG equation can be split into 
a background component and a linearized perturbation component, as follows:
\begin{IEEEeqnarray}{rl}
    & \ddot{\bar{\phi}} + 3H\,\dot{\bar{\phi}} 
    + m^2\,\bar{\phi} = 0, \label{eq.2.1-4}\\
    & \delta\ddot{\phi} +3 H\,\delta\dot{\phi} + 
    \left(\frac{k^2}{a^2}+m^2\right)\delta\phi 
    -2 \Psi\,(\ddot{\bar{\phi}} +3H\,\dot{\bar{\phi}}) \nonumber\\
    & -\left(\dot{\Psi}+3\dot{\Phi}+\frac{k^2}{a}B\right)\dot{\bar{\phi}} = 0, \label{eq.2.1-5}
\end{IEEEeqnarray}
where the $k$ is the comoving wavenumber of a Fourier mode,
$H={\dot{a}}/{a}$ is the Hubble parameter
and the overdot denotes the derivative with respect to the coordinate time $t$.

In this paper, we work with the hydrodynamic representation
rather than the field representation.
We introduce the hydrodynamic variables for the complex SFDM in \cref{ssec:hydrovar}.
Then, we present two novel ingredients of the formalism
uniquely associated with the complex scalar field model:
a particle-based fluid velocity and the stiff gauge condition,
in \cref{ssec:eckart} and \cref{ssec:stiffgauge}, respectively.

\subsection{Hydrodynamic variables}\label{ssec:hydrovar}

We now derive the (linearized) hydrodynamics of the complex SFDM 
based on its stress-energy tensor.
As a result of the least action principle,
the stress-energy tensor of the complex scalar field described by \cref{eq.2.1-1}
is given by
\begin{equation}\label{eq.2.1-7}
    T_{\mu\nu}= \partial_{\mu}\phi\,\partial_{\nu}\phi^*
    +\partial_{\mu}\phi^*\,\partial_{\nu}\phi - g_{\mu\nu}
    \left(g^{\alpha\beta}\,\partial_{\alpha}\,\phi\partial_{\beta}\phi^* -  m^2\,\phi^*\phi\right).
\end{equation}

The above expression implies that unlike the real SFDM 
or ultralight axions, the complex SFDM is an \emph{imperfect} fluid.
This is one of the key messages of the paper.
To see it, we start with the decomposition of the stress-energy tensor 
of a generally imperfect fluid (e.g.,~\cite{2013rehy.book.....R}),
\begin{equation}\label{eq.2.1-6}
  T_{\mu\nu} =  (\rho+P)\,u_\mu u_\nu - g_{\mu\nu}\,P+u_\mu q_\nu+u_\nu q_\mu - \pi_{\mu\nu},
\end{equation}
where $u^\mu$ is the 4-velocity field of an arbitrary family of observers,\footnote{
Such observers are not related to those which define the coordinate system 
or the gauge of the perturbed spacetime, 
but rather (fictitious) observers of the fluid.
Their velocities can be interpreted as the velocity of the fluid.
}
$\rho$ and $P$ are the proper energy density and pressure of the fluid 
measured by these observers in their rest frames,
$q^\mu$ is the energy flux and $\pi_{\mu\nu}$ is the anisotropic stress due to viscosity effects.
$q^\mu$ and $\pi_{\mu\nu}$ characterize the deviation from a perfect fluid.
For a perfect fluid, there is a preferred family of ``comoving observers''
such that both $q^\mu$ and $\pi_{\mu\nu}$ vanish in their rest frames.
However, such comoving observers cannot be found for the complex SFDM, 
given the expression of the stress-energy tensor in \cref{eq.2.1-7}.
An imperfect-fluid description is hence necessary;
more discussions on the imperfect fluid nature of complex SFDM
can be found at the end of \cref{app:hydrovar}.

To proceed with the linear theory, we write the 4-velocity of the observers
as $u^\mu=\left(1-\Psi, \vec v/a\right)$, $|\vec v|\ll c$.
In addition, $q^\mu$ and $\pi_{\mu\nu}$ are also linear perturbations.
Since $u_\mu q^\mu=u^\mu\pi_{\mu\nu}=0$ (see eq.~[\ref{eq.A-1}]),
it can be shown that both $q^0$ and $\pi_{0\mu}$ vanish in linear theory.
Thus, we only need to deal with their spatial components,
for which we can define the following 3-vector and 3-tensor 
on const-$t$ hypersurfaces:
$Q^i\equiv a\,q^i$, $\Pi_{ij}\equiv-\pi_{ij}/a^2$.
The indices of $v^i$, $Q^i$ and $\Pi_{ij}$
are raised and lowered using the conformal background metric, $\delta_{ij}$. 
It can then be shown that in linear theory,
\begin{IEEEeqnarray}{rCl}
  T^0_{~0} & = & \rho \equiv \bar{\rho} + \delta\rho, \label{eq.2.1-10}\\
  T^i_{~0} & = & (\bar{\rho}+\bar{P})\,v^i/a + Q^i/a, \label{eq.2.1-11}\\
  T^{i}_{~j} & = & -P\,\delta^i_{~j} - \Pi^i_{~j}
  \equiv -(\bar{P}+\delta P)\,\delta^i_{~j} - \Pi^i_{~j}, \label{eq.2.1-12}
\end{IEEEeqnarray}
where we define the density and pressure perturbations.
For a complex scalar field, 
we can show that its anisotropic stress vanishes to linear order, $\Pi^i_{~j}=0$
(see \cref{app:hydrovar}).
More relationships between these hydrodynamic variables
and the field are presented in \cref{app:hydrovar}.

The hydrodynamic equations follow from 
the conservation of the stress-energy tensor, $\nabla_\mu T^\mu_{~\nu} = 0$.
The continuity equation for the homogeneous background reads
\begin{equation}\label{eq.2.1-13}
  \dot{\bar{\rho}} + 3H\,(1+w)\,\bar{\rho} = 0,
\end{equation}
where $w\equiv\bar{P}/\bar{\rho}$ is the equation-of-state parameter
of the complex SFDM.
The continuity equation and the Euler equation for the perturbations are written as
\begin{IEEEeqnarray}{rCl}
    \dot{\delta} & = & - 3H\left( \frac{\delta P}{\delta\rho} -w\right)\delta 
    +(1+w)\left(\frac{k^2}{a} v + 3\dot{\Phi}\right) 
    +\frac{k^2}{a}\frac{q}{\bar{\rho}},\qquad \label{eq.2.1-14}\\
    \dot{v} & = & \dot{B} -H(1-3w)\,(v-B) - \frac{\dot{w}}{1+w}(v-B) \nonumber\\
    & & - \frac{1}{a(1+w)}\,\frac{\delta P}{\bar\rho} - \frac{1}{a}\Psi 
    - \frac{1}{1+w} \frac{\dot{q}+4Hq}{\bar{\rho}}, \label{eq.2.1-15}
\end{IEEEeqnarray}
where $\delta\equiv\delta \rho/ \bar{\rho}$ is the density contrast, 
$v\equiv -i \,k_i v^i/k^2 $ and $q\equiv -i\,k_i Q^i/k^2$ 
are the scalar potentials for the 3-velocity and the 3-energy flux, respectively.

The fluid equations (\ref{eq.2.1-14}-\ref{eq.2.1-15}) still contain
ambiguities arising from (i) the gauge freedom and, more importantly, 
(ii) the undetermined velocities of the observers, $u^\mu$.
Since there is no unanimously preferred $u^\mu$ for an imperfect fluid,
we must specify our choice for the 4-velocity of the complex SFDM.

\subsection{The U(1) charge and the Eckart frame}\label{ssec:eckart}

To fix the 4-velocity of a cosmic fluid, one typically has two choices:
the Landau-Lifshitz frame \cite{1959flme.book.....L} 
and the Eckart frame \cite{1940PhRv...58..919E}. 
In the Landau-Lifshitz frame, the energy flux is required to vanish, $q^\mu = 0$, 
so that the 4-velocity must be an eigenvector of the stress-energy tensor, 
i.e., $T^{\mu}_{~\nu}\,u^\nu=\rho\,u^\mu$. 
However, solving for the eigenvalues of
the stress-energy tensor in \cref{eq.2.1-7}
yields extremely complicated expressions
that are impractical to work with algebraically.
Therefore, we adopt the Eckart frame for the complex SFDM,
which is related to the transport of particles, as discussed below. 

For a complex scalar field described by \cref{eq.2.1-1}, 
the difference between its number of particles
and its number of antiparticles is conserved. 
This is the Noether charge associated with the U(1) symmetry of the Lagrangian.
To be precise, the Noether 4-current is defined as
\begin{equation}\label{eq.2.1-16}
  J_\mu = i\left( \phi^*\,\partial_{\mu}\phi-\phi\,\partial_{\mu}\phi^* \right),
\end{equation}
and its conservation law is written as $\nabla_\mu J^\mu=0$.

The Eckart frame for complex SFDM 
is then defined by the 4-velocity associated with the above Noether current,
\begin{equation}\label{eq:fourvelocity}
    u^\mu\equiv {J^\mu}/{n},\qquad n\equiv \pm\sqrt{J_\mu J^\mu},
\end{equation}
where the sign of $n$ is determined 
such that $u^\mu$ is a 4-velocity (future-directed). 
This $u^\mu$ is interpreted as the streaming velocity of the charge,
at which a comoving observer sees no net charge flux.
In \cref{eq:fourvelocity}, $n$ is interpreted as the proper charge density.
In other words, $J^\mu|_E = (n, 0)$ in the locally-inertial rest frames
of the comoving observers that define the Eckart frame.
The proper charge/number density is a new hydrodynamic variable
for ultralight dark matter models, a unique signature of complex SFDM.

We can linearize the charge density by
\begin{equation}\label{eq.2.1-18}
      J^0 = (1-\Psi)\,n \equiv (1-\Psi)\,\bar{n} + \delta n.
\end{equation}
The relationships between the charge-related hydrodynamic variables
and the scalar field 
are provided in \cref{app:hydrovar}. 

The conservation law above yields the EoM for the proper charge density.
The homogeneous evolution is again specified by the continuity equation,
\begin{equation}\label{eq.2.1-19}
  \dot{\bar{n}} + 3H\,\bar{n} = 0.
\end{equation}
Following \cite{2014PhRvD..89h3536L},
we define the conserved comoving charge density for the background field as
\begin{equation}\label{eq.2.1-21}
  N \equiv \bar{n} a^{3}
  = i\,(\bar{\phi}^* \dot{\bar{\phi}} - \bar{\phi} \dot{\bar{\phi}}^*)\,a^3,
\end{equation}
which does not vanish in general. 
In fact, $N$ is an independent parameter for the complex scalar field model,
distinguished from the singlet, real field model (such as axions).
When $N\simeq\rho_\mathrm{dm}\,a^3/m$ at late times, 
the background field rotates in the complex plane
at an angular velocity $\omega=m$ \cite{2014PhRvD..89h3536L}, 
known as the spintessence limit \cite{2002PhLB..545...17B}. 
When $N = 0$, the complex field is effectively reduced to a real field
(the real-field limit) \cite{2002PhRvD..65h3514A}.

The perturbation equation for the proper charge density also 
follows from its conservation law, $\nabla_\mu J^\mu=0$, such that
\begin{equation}\label{eq.2.1-20}
  \dot{\delta}_n = \frac{k^2}{a}\, v + 3\dot{\Phi},
  \qquad \delta_n \equiv\delta n /\bar{n}.
\end{equation}

\subsection{Constitutive equations and the stiff gauge}\label{ssec:stiffgauge}

The above EoMs for the perturbation variables, 
eqs.~(\ref{eq.2.1-14}), (\ref{eq.2.1-15}) and (\ref{eq.2.1-20}),
are not sufficient to close the system.
Typically, additional relations between the stress and the density 
is needed, known as the \emph{constitutive equations}.
In this case, the scalar-mode constitutive equations for complex SFDM
are given by $\delta P\,(\delta\rho,\,\delta n)$ and $q\,(\delta\rho,\,\delta n)$,
which we will derive here.

Following the prescription in~\cite{1998ApJ...506..485H}, 
the relationship between $\delta P$ and $\delta\rho$ 
is first derived under a particular coordinate/gauge condition
and then transformed to any general coordinate system.
For the complex scalar field, we impose the condition that
$\bar{\phi}\,\delta \phi^* + \bar{\phi}^*\,\delta \phi=0$.
It yields
\begin{equation}\label{eq.3.1-1}
    c_s^2\equiv \left(\frac{\delta P}{\delta\rho} \right)\Bigg|_s=1,
\end{equation}
where the subscript $s$ denotes this particular gauge condition,
which we call the \emph{stiff gauge} from now on,
since the sound speed under this condition, $c_s$, reaches the speed of light.
Note that the stiff gauge coincides with the comoving gauge ($v=B$)
and the uniform-field gauge ($\delta\phi=0$) for a real scalar field, 
but these conditions are not equivalent to each other for a complex field.
Transforming \cref{eq.3.1-1} to arbitrary coordinates, we obtain
\begin{equation}\label{eq.3.1-3}
   \delta P =  \delta\rho +( \dot{\bar{\rho}}-\dot{\bar{P}})\,a (v - B) -(\dot{\bar{\rho}}-\dot{\bar{P}})\,a (v - B)|_s.
\end{equation}
The value of $(v - B)|_s$ in the stiff gauge can be further expressed 
in terms of the hydrodynamic variables, as derived in \cref{app:identities}.
Eliminating $(v - B)|_s$, we rewrite \cref{eq.3.1-3} as
\begin{equation}\label{eq.3.1-10}
  \begin{split}
    \delta P = &~\delta\rho 
    -3\left(1-c_\mathrm{ad}^2\right)\,H\,\bar{\rho}\,(1+w)\,a (v - B)  \\
      & -\bar{\rho}\,(1-w)\left[\delta_n + \Psi 
      - 2 aH\,(v - B) + a (\dot{v} - \dot{B}) \right],
  \end{split}
\end{equation}
where $c_\mathrm{ad}^2\equiv \ud \bar P/\ud\bar\rho
={\dot{\bar P}}/{\dot{\bar\rho}} = w - {\dot{w}}/[{3H(1+w)}]$ 
is the adiabatic sound speed.

The second constitutive equation involves the energy flux, $q$,
particular to the complex SFDM model.
Derived in \cref{app:identities}, it is expressed as
\begin{IEEEeqnarray}{rCl} \label{eq.3.1-11}
  q & = & -\frac{1}{4m^2} \frac{(\dot{\bar{\rho}}-\dot{\bar{P}})^2}{\bar{\rho}-\bar{P}} (v-B)|_s \nonumber\\
  & = & \frac{3\left(1-c_\mathrm{ad}^2\right)}{4m^2\,a}\,H\bar{\rho}\,(1+w)\, \nonumber\\
  & & \cdot\left[{\delta_n} +\Psi 
  - 2aH\,(v - B) +a (\dot{v} -\dot{B}) \right].\qquad 
\end{IEEEeqnarray} 
Thus, the energy flux is proportional to the charge density perturbation.

Eqs.~(\ref{eq.3.1-10}) and (\ref{eq.3.1-11}) provide
the desired two scalar-mode constitutive equations. 
They are part of the main results of this paper.
The EoMs of the complex SFDM will be numerically solved along with
these constitutive equations as well as the standard Einstein equations.

\section{Background evolution}\label{section4}

Our results concerning the background evolution of the complex SFDM 
are discussed in this section. 
In our model, the complex SFDM makes up the total dark matter sector.
All other sectors are the same as those in $\Lambda$CDM.
We adopt the \emph{Planck} 2018 best-fit cosmological parameters 
\cite{2020A&A...641A...6P} for all the numerical cases shown in this paper.

Before presenting the numerical solutions, we first give a brief review of
the analytical behaviors of the homogeneous evolution.
For most of the expansion history,
we can assume a power-law relationship
between the scale factor and the cosmic time, $a \propto t^p$. 
Eq.~(\ref{eq.2.1-4}) can then be rewritten as

\begin{figure}[H]
  \centering
  \includegraphics[width=7.5cm]{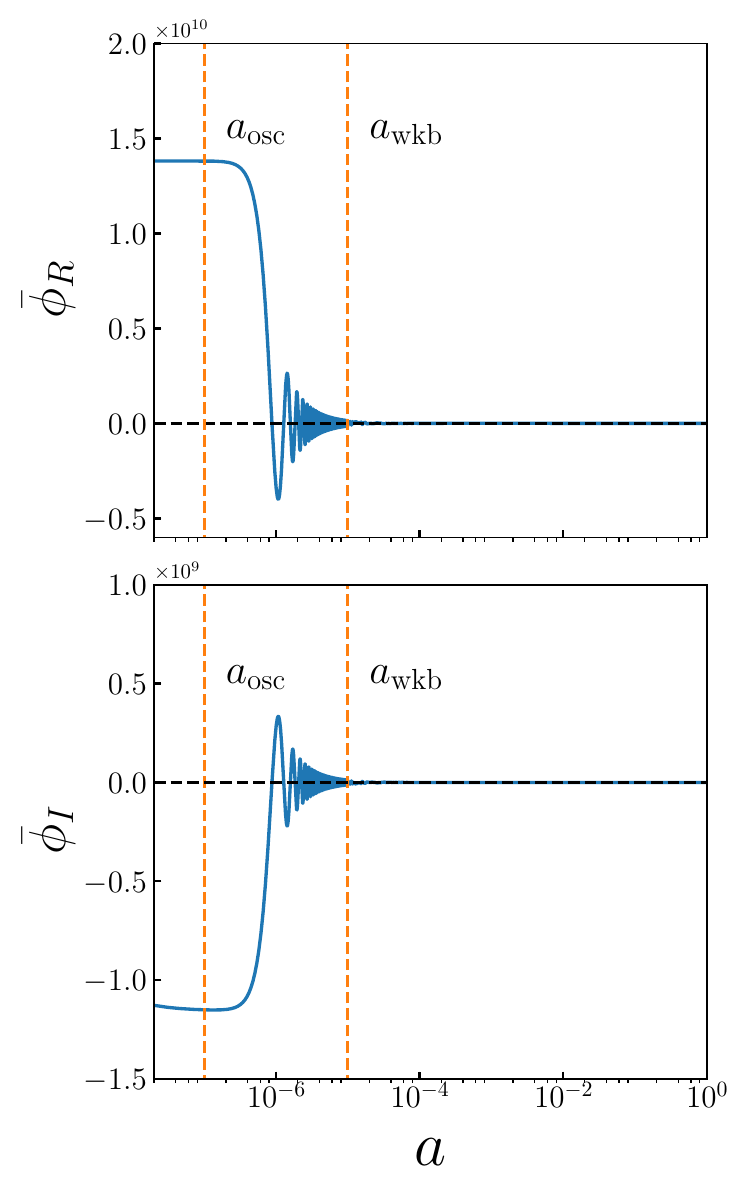}
  \caption{Example evolution of the homogeneous complex scalar field.
   In this case, $m = 10^{-22}\,\mathrm{eV}$, 
   $\dot{R}_{\mathrm{ini}} = 10^{7}\,\mathrm{eV}^2$
   and $N = 10^{10}\,\mathrm{cm}^{-3}$.
  The upper panel shows the real part of the complex field
  and the lower panel shows the imaginary part. 
  When $a \sim a_{\mathrm{osc}}$, 
  the oscillation regime of the complex SFDM begins. 
  We solve the KG equation exactly until $a=a_{\mathrm{wkb}}$,
  after which we switch to the WKB approximation.}\label{Fg.4-1}
\end{figure}

\begin{equation}\label{eq.4.1-1}
  \ddot{\bar\phi}+3\frac{p}{t}\,\dot{\bar\phi}+m^2\,\bar\phi=0.
\end{equation}
Its analytical solutions are expressed in terms of Hankel functions
\cite{2010PhRvD..82j3528M},
\begin{equation}\label{eq.4.1-2}
  \bar\phi(t)  =  a^{-3/2}\left(\frac{t}{t_0}\right)^{1/2} \left[C_1 H^{(1)}_n \left(m t\right)+C_2 H^{(2)}_n\left(m t\right) \right],
\end{equation}
where $n= (3p-1)/2$, $H^{(1)}_n(x)$ and $H^{(2)}_n(x)$ are 
the Hankel functions of the first and second kind, respectively.
The well-known oscillation regime of a (real) scalar field in which $H\ll m$ 
(e.g., \cite{1983PhRvD..28.1243T}) results from 
the late-time asymptotic behavior of Hankel functions, such that
\begin{equation}\label{eq.4.1-3}
    \bar\phi(t) \approx a^{-3/2}\sqrt{\frac{1}{2m}}\,
    \left(\psi_{+}\,e^{-imt}+\psi_{-}\,e^{imt}\right),
\end{equation}
where $\psi_{+}$ and $\psi_{-}$ are constant complex coefficients.
This solution is essentially the WKB approximation
\cite{2009PhLB..680....1H, 2013PhLB..726..559N},
in which a fast-oscillating part can be separated from a slowly evolving part.
In general, even when the $a \propto t^p$ assumption is not applicable, 
the above asymptotic solution holds for $H\ll m$.
For a complex scalar field, the concept of the oscillation regime
is actually extended and refers to
the generic late-time \emph{rotational} behavior of the field in the complex plane,
covering both the spintessence limit \cite{2002PhLB..545...17B, 2014PhRvD..89h3536L}
and the oscillatory, real-field limit \cite{2002PhRvD..65h3514A}.

\begin{figure}[H]
  \centering
  \includegraphics[width=7.5cm]{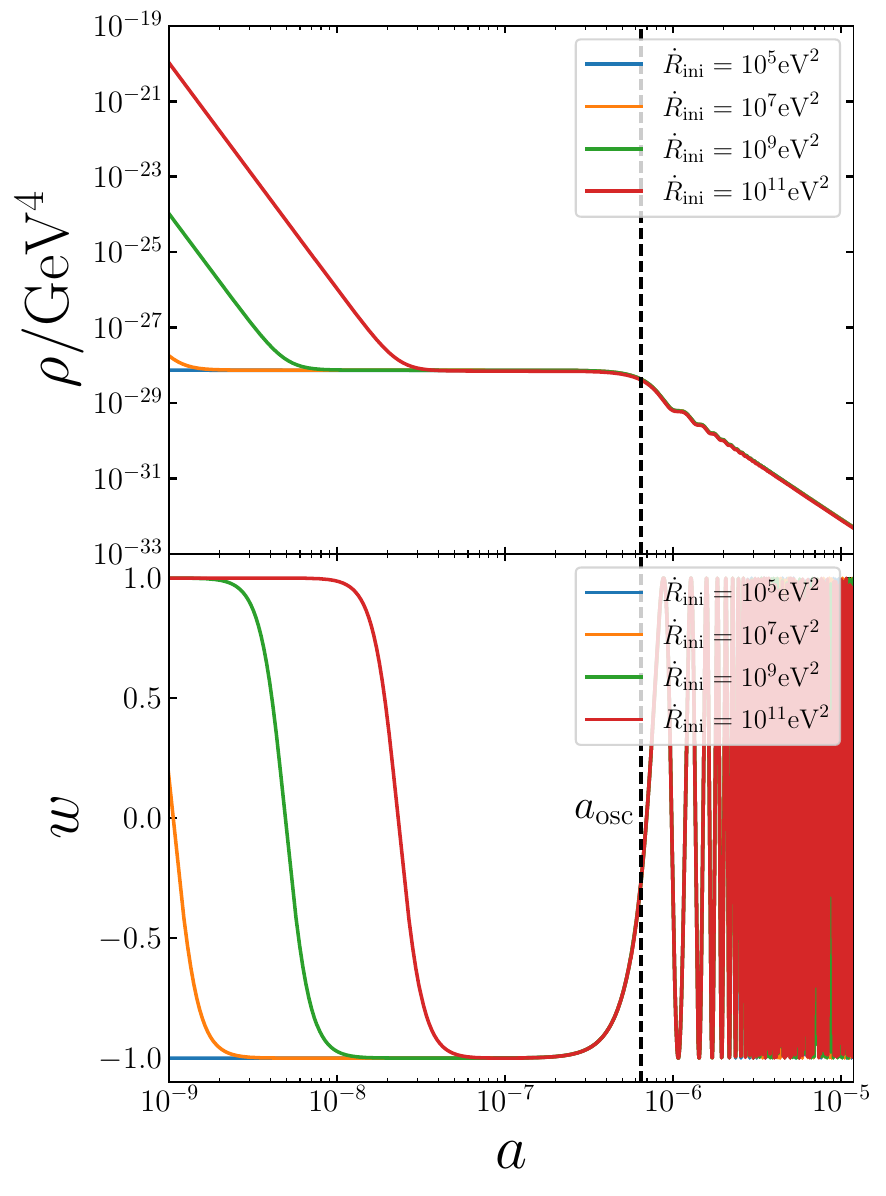}
  \caption{Illustrative cases of the evolution of
  the homogeneous energy density $\bar{\rho}$ of the complex SFDM (the upper panel)
  and its equation-of-state parameter $w$ (the lower panel),
  with various initial conditions $\dot{R}_{\mathrm{ini}}$. 
  Here we adopt $m = 10^{-22}\,\mathrm{eV}$
  and $ N = 10^{10}\,\mathrm{cm}^{-3}$.
  The lower panel shows the stiff/kination phase ($w=1$)
  as the earliest stage of the homogeneous evolution
  as well as an intermediate slow-roll phase ($w=-1$).
  The duration of the $w=-1$ phase
  is mainly dependent on the value of the parameter $\dot{R}_{\mathrm{ini}}$.}\label{Fg.4-2}
\end{figure}

Since the oscillation/rotation period scales as $m^{-1}$, 
it is impractical to track the exact evolution of the complex SFDM numerically
for a long duration of time throughout the history of the universe.
This challenge is often circumvented by switching to
the time-averaged solution when the oscillations begin
(the slowly evolving coefficients in the WKB approximation)
\cite{2014PhRvD..89h3536L, 2015PhRvD..91j3512H, 2020PhRvD.101b3501C,2022PhRvD.105l3529P}.
Following \cite{2015PhRvD..91j3512H}, we denote the approximate scale factor
at the onset of oscillations as $a_\mathrm{osc}$, 
defined by $H(a_\mathrm{osc})=3m$.
In this work, we solve the KG equation exactly 
until a chosen moment, $a = a_\mathrm{wkb}$, 
after which we switch to the WKB approximation. 
We require that $a_\mathrm{wkb}/a_\mathrm{osc}>\mathcal{O}(10)$.
To describe the initial conditions of the background field, 
we rewrite the complex scalar field in terms of its amplitude and phase,
$\bar{\phi} = R e^{i\theta}$. $\dot{R}$ is the time derivative of $R$, 
which represents the rate of change of the field amplitude
(also referred to as the radial velocity, while $\dot{\theta}$ is known as the angular velocity).
Overall, the background evolution of the complex SFDM is determined
by the following set of parameters:
$\{m, \Omega_{\phi}h^2, N, \dot{R}_{\mathrm{ini}}\}$,
where $\dot{R}_{\mathrm{ini}}$ is the value of
$\dot{R}$ at the initial moment $a_{\mathrm{ini}}$.
More details of the numerical algorithm are described in \cref{app:numerical}.

\cref{Fg.4-1} shows an example solution, 
where the real part and the imaginary part of the background field
are plotted separately, $\bar\phi\equiv(\bar\phi_R+i\,\bar\phi_I)/\sqrt{2}$. 
In the oscillation regime,
the complex SFDM is known to behave as a pressureless dust like CDM
\cite{2014PhRvD..89h3536L}.
Its energy density and pressure are well captured by the time-averaged solution, 
for which $\langle \bar{\rho}\rangle\propto a^{-3}$,
$\langle \bar{w}\rangle=\langle \bar{P}/\bar\rho\rangle\approx 0$ 
\cite{2014PhRvD..89h3536L, 2017PhRvD..96f3505L}
and $\langle c_\mathrm{ad}^2\rangle\approx 0$,
as illustrated in \cref{Fg.4-2}.
The lower panel of \cref{Fg.4-2} shows the evolution
of the equation-of-state parameter of the complex SFDM
for various initial conditions $\dot{R}_{\mathrm{ini}}$. 
The value of $\dot{R}_{\mathrm{ini}}$ sets
the initial kinetic energy of the complex scalar field.
When it is large enough, the energy density of the field
can initially be dominated by the kinetic energy, 
and thus the scalar field starts in the stiff phase ($w = 1$), 
also known as ``kination''
\cite{1993PhLB..315...40S, 1997PhRvD..55.1875J, 2014PhRvD..89h3536L, 2020PhRvL.124y1802C, 2021JCAP...10..024L,
2025ApJ...985..117L, 2025arXiv250404054Z}.
As \cref{Fg.4-2} shows, the kinetic energy decays quickly
due to the Hubble friction,
and the scalar field enters the slow-roll phase ($w=-1$)
dominated by its potential energy \cite{2002PhRvD..65h3514A, 2016PhR...643....1M}.
The slow-roll phase ends when the field begins to oscillate
(around $a_\mathrm{osc}$).
Therefore, the duration of this intermediate slow-roll phase
is dependent on both $\dot{R}_{\mathrm{ini}}$ and $m$, the boson mass.
For higher values of $\dot R_\mathrm{ini}$ and $m$,
the duration of the $w=-1$ phase is shorter and may eventually vanish.
All these behaviors are similar to those of a real scalar field.

\section{Perturbation evolution of complex SFDM}\label{section5}

In this section, we present the results of our perturbation calculation.
We update the hydrodynamic equations for perturbations
using variables that match those integrated by the {\footnotesize CAMB} code
in \cref{sec:perturbeq}. 
We specify the initial conditions and discuss the numerical solutions
in \cref{sec:ICsol}.

\subsection{Dimensionless perturbation equations}\label{sec:perturbeq}

For pedagogical purposes,
we choose the conformal Newtonian gauge ($B=E=0$)
to derive the perturbation equations.
The perturbed FLRW metric is thus written as
\begin{equation}\label{eq.5.1-1}
  \ud s^2 = a^2 \left[(1 + 2\Psi_A)\,\ud\eta^2 
  - (1-2\Phi_H)\,\delta_{ij}\,\ud x^i \ud x^j \right],
\end{equation}
where $\ud\eta = \ud t/a$ is the conformal time,
$\Psi_A$ and $\Phi_H$ are the gauge-invariant Bardeen potentials \cite{1980PhRvD..22.1882B}.
We define the Fourier-space dimensionless momentum density as 
\begin{equation}
    \upsilon \equiv -k\,(1+w)(v-B),
\end{equation}
and the dimensionless energy flux as $\epsilon \equiv -kq/ \bar{\rho}$.
We further refer to the sum, $\upsilon+\epsilon$,
as the dimensionless heat flux.\footnote{It is worth clarifying
a possible confusion about the usage of the term ``heat flux'' in the literature:
the momentum density $\upsilon$ has been referred to as ``heat flux'' 
in {\footnotesize CAMB} \cite{Lewis:1999bs} (\url{https://cosmologist.info/notes/CAMB.pdf})
and {\footnotesize axionCAMB} \cite{2015PhRvD..91j3512H}.
However, what appears in the equations is
actually the sum $\upsilon+\epsilon$ rather than $\upsilon$ alone;
the former is what we mean by heat flux in this paper.
Only when $\epsilon$ vanishes can the momentum density, $\upsilon$, 
be identified with the heat flux here
(which is true for the ordinary cosmic components
considered by {\footnotesize CAMB} and {\footnotesize axionCAMB}).}

Given these dimensionless variables, 
eqs.~(\ref{eq.2.1-14}), (\ref{eq.2.1-15}) and (\ref{eq.2.1-20}) can be recast into
\begin{IEEEeqnarray}{rCl}
    \delta' & = & 3\mathcal{H}w\,\delta-3\mathcal{H}\,\frac{\delta P}{\bar\rho}
    -k\,\upsilon +3(1+w)\,\Phi_H' - k\,\epsilon, \qquad \label{eq.5.1-2} \\
    \upsilon' & = & -\mathcal{H}\,(1-3w)\,\upsilon + k\,\frac{\delta P}{\bar\rho}
    + (1+w) \,k\,\Psi_A \nonumber\\
    & & -\,\epsilon' - \mathcal{H}(1-3w)\,\epsilon, \label{eq.5.1-3} \\
    \delta_n' & = & -\frac{1}{1+w}\,k\,\upsilon + 3 \Phi_H', \label{eq.5.1-4}
\end{IEEEeqnarray}
where the prime denotes the derivative with respect to the conformal time 
and $\mathcal{H} \equiv a'/a$ is conformal Hubble rate. 
Note that eqs.~(\ref{eq.5.1-2}) and (\ref{eq.5.1-3}) reduce to 
eqs.~(20) and (21) in Ref.~\cite{2015PhRvD..91j3512H} 
in the real-field, ultralight axion case.\footnote{We have taken into account
the gauge transformation between the Newtonian gauge used here 
and the synchronous gauge used in Ref.~\cite{2015PhRvD..91j3512H}.}
The dimensionless Euler equation (\ref{eq.5.1-3}) exhibits the symmetry between
the dissipationless, momentum density ($\upsilon$)
and the dissipative energy flux ($\epsilon$).
The presence of the energy flux is a signature of the complex SFDM model.
The total energy transport for complex SFDM is described by
the heat flux, $\upsilon+\epsilon$.

We can also non-dimensionalize the constitutive equations.
In fact, rearranging eq.~(\ref{eq.3.1-11}) yields 
another ordinary differential equation (ODE) for the momentum density
that does not depend on $\epsilon'$, 
\begin{IEEEeqnarray}{rCl}\label{eq:momentumdensity}
    \upsilon' & = & -\mathcal{H}\,(1-3w)\,\upsilon +3\mathcal{H}\left(1-c_\mathrm{ad}^2\right)\upsilon \nonumber\\
    & & +(1+w)\,k\,(\delta_n+\Psi_A) + \frac{4m^2a^2}{3\mathcal{H}
    \left(1-c_\mathrm{ad}^2\right)}\epsilon.
\end{IEEEeqnarray}
Recall that $c_\mathrm{ad}^2 = w - {{w}'}/[{3\mathcal{H}(1+w)}]$.
Meanwhile, inserting eq.~(\ref{eq.3.1-11}) into eq.~(\ref{eq.3.1-10}) yields
\begin{equation}\label{eq.5.1-5}
    \frac{\delta P}{\bar{\rho}} = \delta
    +3\mathcal{H}\left(1-c_\mathrm{ad}^2\right)\frac{\upsilon}{k}
    +\frac{1-w}{1+w}\,\frac{4m^2a^2}{3 k\,\mathcal{H}\left(1-c_\mathrm{ad}^2\right)}\epsilon.
\end{equation}
Compared with the axion case, 
the above effective equation of state for the perturbed SFDM
contains an extra term (the last term on the r.h.s.)
due to the presence of the dissipative energy flux.
Finally, inserting the above two equations into eq.~(\ref{eq.5.1-3}) 
yields an ODE for the energy flux itself,
\begin{IEEEeqnarray}{rCl}
    \epsilon' & = & - \mathcal{H}(1-3w)\,\epsilon
    +k\,\delta-(1+w)\,k\,\delta_n \nonumber\\
    & & -\frac{2w}{1+w}\,\frac{4m^2a^2}{3\mathcal{H}
    \left(1-c_\mathrm{ad}^2\right)}\epsilon. \label{eq.5.1-6}
\end{IEEEeqnarray}

Finally, we take into account the contribution of the dissipative energy flux to the stress-energy tensor in the Einstein equations.
The linearized Einstein equations read
\begin{IEEEeqnarray}{rl}
  & -k^2 \Phi_H - 3\mathcal{H}(\Phi_H^{\prime} + \mathcal{H}\Psi_A) = 4\pi G a^2 \sum_{x}\delta \rho_x, \qquad \label{eq.E1} \\
  & - (\Phi_H^{\prime} + \mathcal{H} \Psi_A) = 4\pi G a^2 \sum_{x}
  \left(\upsilon_x + \epsilon_x\right),  \qquad    \label{eq.E2} \\
  &  k^2 \left(\Phi_H-\Psi_A\right) 
  = 8\pi G a^2 \sum_{x} \Pi_{x},  \label{eq.E3} \\
  & \Phi_A^{\prime\prime} + \mathcal{H}\Psi_A' + 2\mathcal{H} \Phi' -\frac{1}{3}k^2 (\Psi_A - \Phi_H) \nonumber\\
  & = 4 \pi G a^2 \sum_{x} \delta P_x, \label{eq.E4}
\end{IEEEeqnarray}
where the sum goes over the cosmic components, 
and the scalar-mode anisotropic stress $\Pi$ in eq.~(\ref{eq.E3})
is defined by $\Pi_{ij} \equiv -({k}_i {k}_j/k^2-\delta_{ij}/3)\,\Pi$
for each component.
Eqs.~(\ref{eq.5.1-2}), (\ref{eq.5.1-4})--(\ref{eq.5.1-6}) 
along with the Einstein equations (\ref{eq.E1}-\ref{eq.E4})
form a closed system of ODEs for linear perturbations in a universe
where the dark matter component consists of complex SFDM. 
Note that the complex SFDM is the only contribution to $\sum_x\epsilon_x$
in eq.~(\ref{eq.E2}).

\subsection{Initial conditions and the early-stage evolution}\label{sec:ICsol}

We now consider the linear evolution of the complex SFDM 
in a realistic universe with all ordinary cosmic components, 
including photons ($\gamma$), neutrinos ($\nu$), baryon ($\mathrm{b}$),
a cosmology constant ($\Lambda$),
complex SFDM ($\phi$), and CDM ($\mathrm{c}$).
We assume $\Omega_\mathrm{c}\to0$.\footnote{The {\scriptsize CAMB} code
adopts the synchronous gauge along with the ``CDM frame''.
Hence, it is always more convenient to retain some portion of CDM.}
We will call this a $\Lambda$SFDM universe.
To proceed with numerical calculations, we use the standard adiabatic initial conditions \cite{1995ApJ...455....7M}.
Particularly, the initial conditions for perturbations in a $\Lambda$SFDM universe
are written as
\begin{equation}\label{eq.5.2-2}
    \begin{split}
        \delta_{\mathrm{ini}} = & -\frac{3}{2}(1+w_{\mathrm{ini}}) \,\Psi_{A,\,\mathrm{ini}}, \\
        \delta_{n,\,\mathrm{ini}} = & -\frac{3}{2}\,\Psi_{A,\,\mathrm{ini}}, \\
        \upsilon_{\mathrm{ini}} = &~\epsilon_{\mathrm{ini}} = 0, \\
        \Psi_{A,\,\mathrm{ini}} =& ~ \frac{20}{15 +4R_\nu}
        \mathcal{R}_\mathrm{prim}, \qquad \Phi_{H,\,\mathrm{ini}} 
        = (1+\frac{2}{5}R_\nu)\,\Psi_{A,\,\mathrm{ini}},
    \end{split}
\end{equation}
where $R_\nu \equiv \bar{\rho}_{\nu}/(\bar{\rho}_\nu + \bar{\rho}_{\gamma})$,
and $\mathcal{R}_\mathrm{prim}$ is the primordial curvature perturbation. 
We add the hydrodynamic perturbation equations of the complex SFDM, 
eqs.~(\ref{eq.5.1-2}), (\ref{eq.5.1-4})--(\ref{eq.5.1-6}), 
to the {\footnotesize CAMB} code.
Accordingly, all the figures we present in this section show
the synchronous-gauge values of the fluid variables (with the subscript `syn'),
following the convention of {\footnotesize CAMB}.
For fixed model parameters and initial conditions, 
we first solve for the background solution as described in \cref{section4},
and then use it as an interpolation table
for the background quantities in the perturbation equations when necessary.

The major computational challenge for integrating the perturbation equations
is again the late-time oscillatory behavior of the scalar field. 
Apart from the oscillations of the homogeneous field
(with an angular frequency of $m$),
the perturbations are themselves oscillatory at late stages of their evolution, 
when $k^2+(ma)^2\gg\mathcal{H}^2$ \cite{2015PhRvD..91j3512H, 2022PhRvD.105l3529P}.
This can be inferred from the perturbed KG equation, eq.~(\ref{eq.2.1-5}).
These high-frequency oscillations make it impossible to
follow the exact time evolution of the ODE system over cosmological time scales.
The typical treatment for the oscillation regime of perturbations
is again to apply the WKB approximation to obtain some cycle-averaged,
``effective fluid'' description \cite{2015PhRvD..91j3512H, 2016JCAP...07..048U,2017PhRvD..95l3511H}.
However, such approximations may induce significant errors
for certain parameter ranges \cite{2020PhRvD.101b3501C,2022PhRvD.105l3529P}.

In this paper, we do not attempt to solve for the full perturbation evolution
of $\Lambda$SFDM, which inevitably involves some late-stage approximations.
Instead, we focus on the early-stage evolution of the perturbations
before numerical integration is prohibited by the rapid oscillations.

\begin{figure}[H]
  \centering
  \includegraphics[width=8cm]{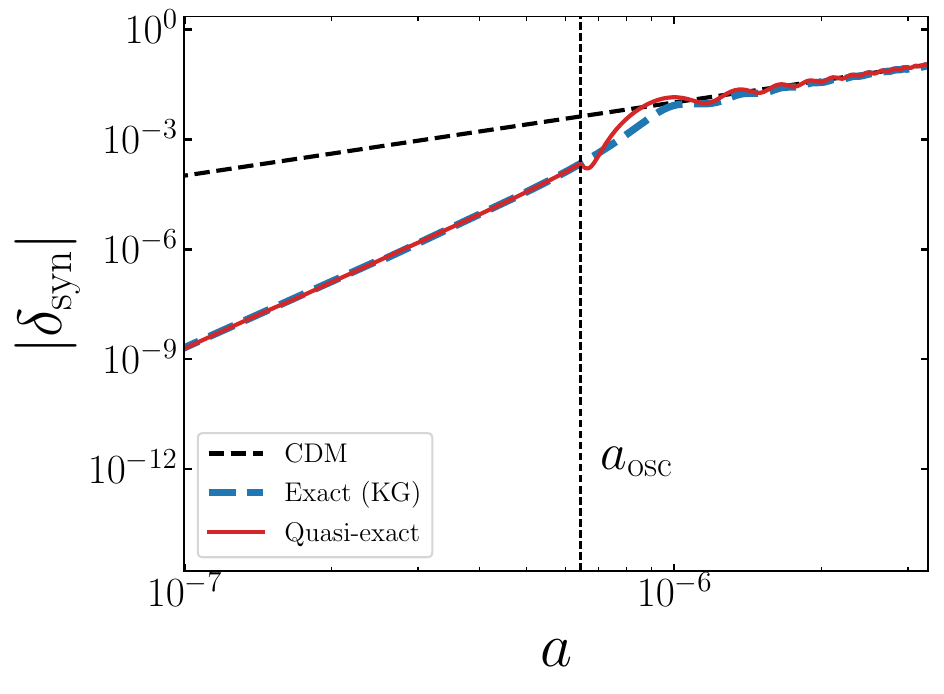}\\
  \caption{Comparison between the exact solution (to the KG equation)
  and the ``quasi-exact'' solution (to the fluid equations)
  in the synchronous gauge.
  For the latter, we set $w=c_{\mathrm{ad}}^2=0$ when $a>a_{\mathrm{osc}}$.
  In this example,
  $m = 10^{-22}\,\mathrm{eV}$, $k = 0.1\,h/\mathrm{Mpc}$, 
  $\dot{R}_{\mathrm{ini}} =10^{5}\,\mathrm{eV}^{2}$ 
  and $N = 10^{10}\,\mathrm{cm}^{-3}$.
  The figure demonstrates the validity of our quasi-exact approximation.
  }\label{Fg.switch}
\end{figure}

For simplicity, we further consider a ``quasi-exact'' approximation
of the exact perturbation solution,
where we take $w=c_{\mathrm{ad}}^2=0$ for $a > a_\mathrm{osc}$. 
Essentially, this quasi-exact treatment assumes that
the switch for the background SFDM quantities
from their early exact evolutions to late-time, oscillation-averaged evolutions
occurs at $a_\mathrm{osc}$,
as far as their effects on the SFDM perturbations
in eqs.~(\ref{eq.5.1-2}), (\ref{eq.5.1-4})--(\ref{eq.5.1-6}) are concerned.
The validity of the quasi-exact approximation is demonstrated in \cref{Fg.switch}. 
It shows that the quasi-exact solution (to the fluid equations)
agrees with the exact solution (to the KG equation),
and both follow the CDM growth function when $m/H\gtrsim 10$.
Meanwhile, the oscillation amplitude of the quasi-exact solution
is reduced compared with that of the exact solution.
In the rest of the paper, we will only show quasi-exact solutions
for illustrations of the early-stage evolution of the complex SFDM.
We will discuss them in detail in what follows.

\begin{figure}[H]
  \centering
  \includegraphics[width=7cm]{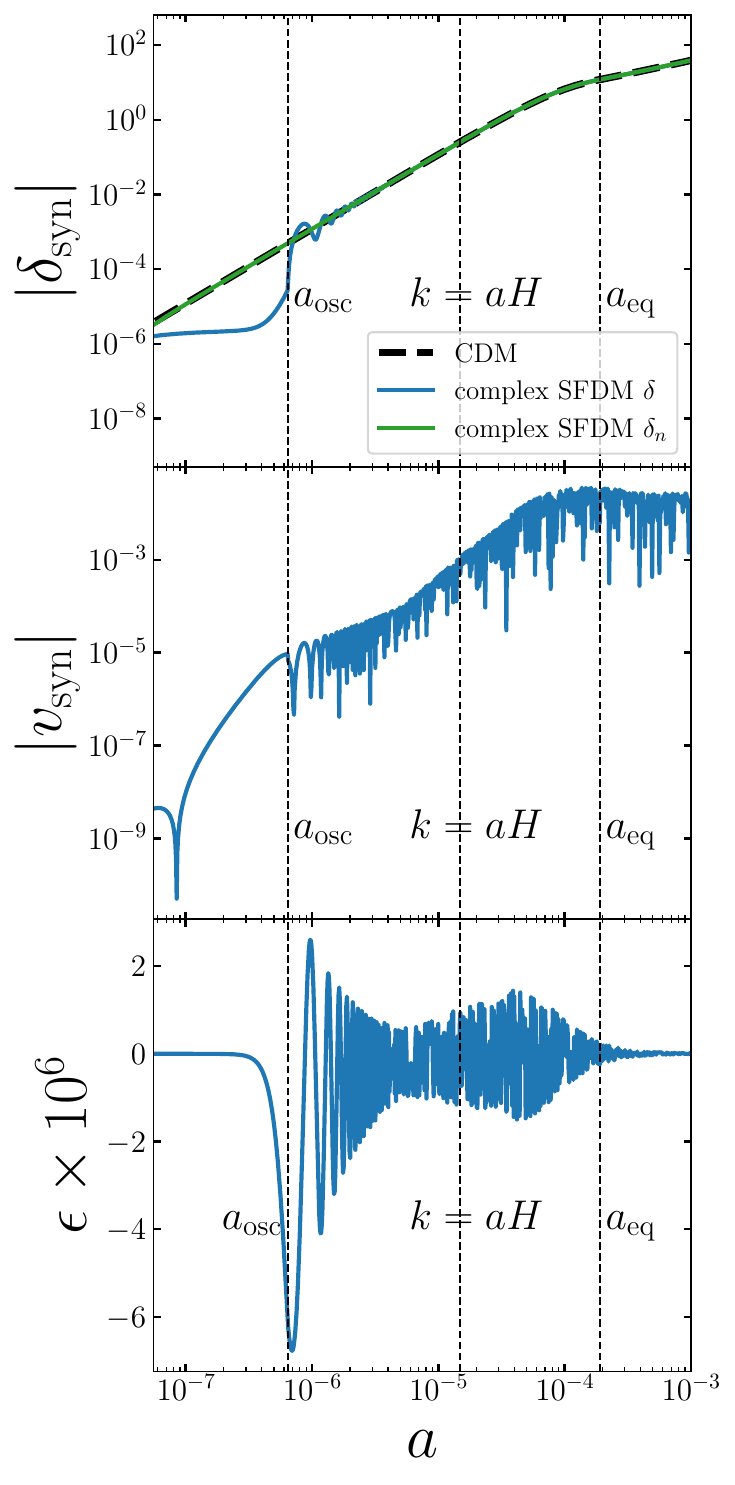}\\
  \caption{Illustrative early-stage evolutions of the SFDM fluid variables 
  in the synchronous gauge.
  In this example, we adopt $m = 10^{-22}\,\mathrm{eV}$, $k = 0.1\,h/\mathrm{Mpc}$,
  $\dot{R}_{\mathrm{ini}} =10^{5}\,\mathrm{eV}^{2}$
  and $N = 10^{10}\,\mathrm{cm}^{-3}$}.\label{Fg.all}
\end{figure}

\cref{Fg.all} displays the evolutions of
the energy density contrast ($\delta$), the charge density contrast ($\delta_n$),
the velocity potential ($v$) and the dissipative energy flux ($\epsilon$)
of the complex SFDM.
The top panel shows that when $a<a_{\mathrm{osc}}$, 
the evolution of the energy density contrast
deviates from that of the charge density contrast.
This is because the latter is only sourced
by the gravity potential (cf. eq.~[\ref{eq.5.1-4}]),
whereas the former is also affected by the pressure term.
Moreover, we note that the evolution of $\delta_n$
closely follows that of the CDM overdensity.
When $a>a_{\mathrm{osc}}$, the SFDM overdensity also catches up with them,
so that $\delta \approx \delta_n$.
The middle panel of \cref{Fg.all} shows the evolution of the velocity potential,
which exhibits a growth trend at all times.
For $a>a_{\mathrm{osc}}$, the velocity potential displays an oscillatory behavior,
even if the effect of the background oscillations on the perturbations
has been factored out in the quasi-exact approximation.
The bottom panel of \cref{Fg.all} reveals that in this example,
the evolution of the dissipative energy flux of the complex SFDM
exhibits an exponential growth for $a<a_{\mathrm{osc}}$.
When $a>a_{\mathrm{osc}}$, it enters an oscillatory phase
so that $\epsilon$ can change its sign.
The amplitude of these oscillations decays over time.
We note that at $a=a_{\mathrm{eq}}$, the energy flux is already close to $0$,
indicating that the energy flux is unimportant in the matter-dominated era.

\begin{figure}[H]
  \centering
  \includegraphics[width=8cm]{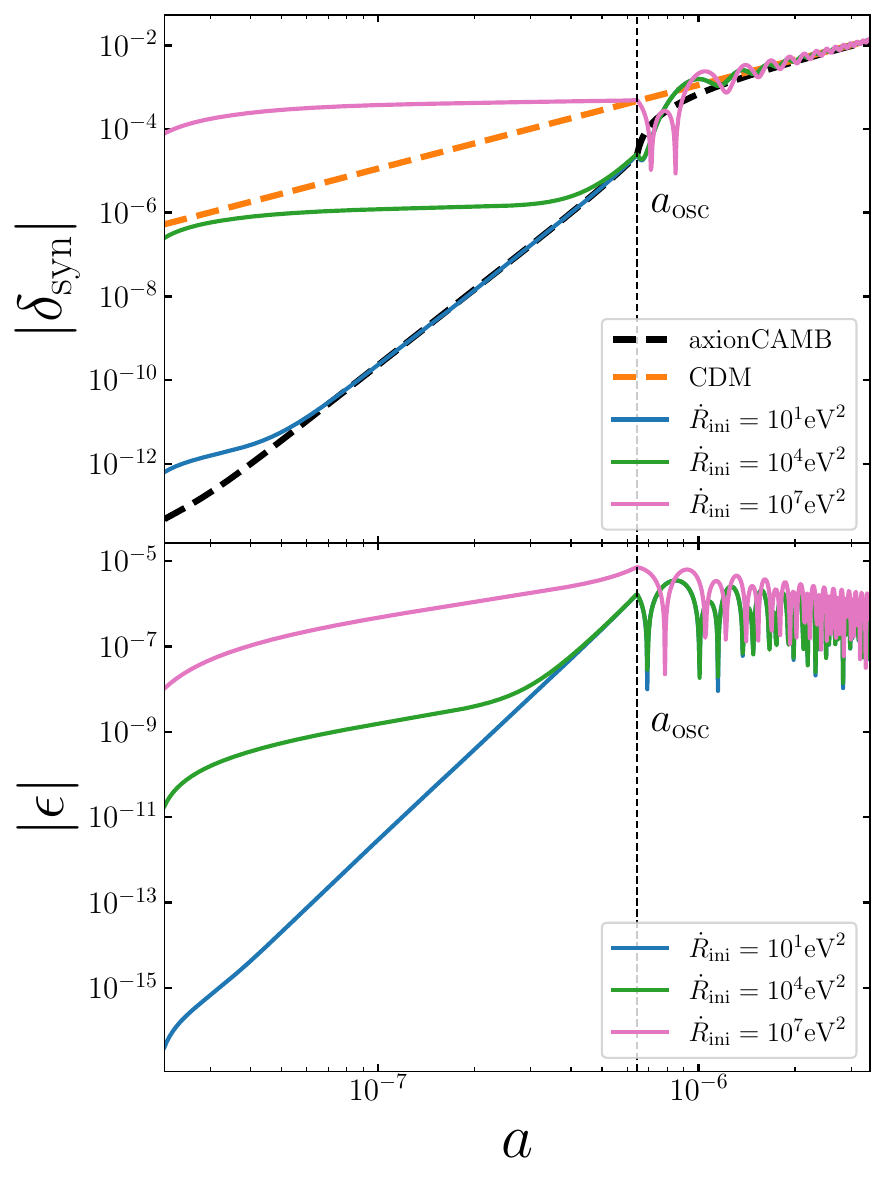}\\
  \caption{\emph{Upper panel}: comparison between the early-stage evolution 
  of the overdensity of the complex SFDM and that of the axion dark matter. 
  \emph{Lower panel}: early-stage evolution of the SFDM energy flux.
  We vary the initial value of $\dot{R}_{\mathrm{ini}}$
  for the background evolution of the complex SFDM.
  The overdensity of the axion dark matter is evolved using axionCAMB.
  Here we adopt $m = 10^{-22}\,\mathrm{eV}$, 
  $N = 10^{10}\mathrm{cm}^{-3}$ and $k = 0.1\,h/\mathrm{Mpc}$.
  }\label{Fg.fordotR}
\end{figure}

\begin{figure}[H]
  \centering
  \includegraphics[width=8cm]{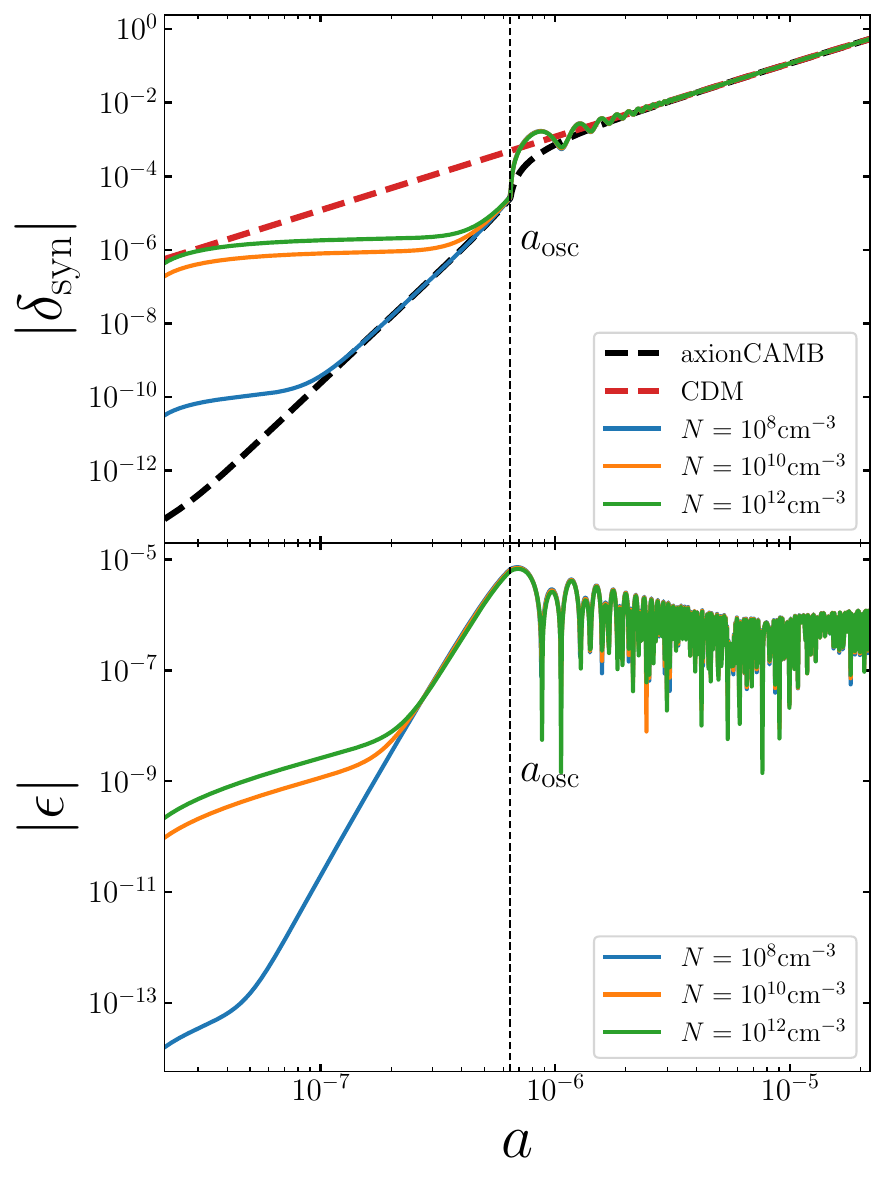}\\
  \caption{Similar to \cref{Fg.fordotR},
  but the value of the conserved charge number, $N$,
  is varied here, instead of $\dot{R}_{\mathrm{ini}}$.
  We adopt $m = 10^{-22}\,\mathrm{eV}$, 
  $\dot{R}_{\mathrm{ini}} = 10^{7}\,\mathrm{eV}^{2}$
  and $k = 0.1\,h/\mathrm{Mpc}$.
  }\label{Fg.forN}
\end{figure}

\cref{Fg.fordotR} and \cref{Fg.forN}
demonstrate that the impact of the energy flux
on the growth of perturbations mainly comes from its early-stage evolution.
In the upper panels, we compare the evolutions of the density contrast
of the complex SFDM (in the presence of the $\epsilon$ term, 
for varying model parameters: $\dot{R}_{\mathrm{ini}}$ or $N$) 
with that of the axion dark matter (in the absence of the $\epsilon$ term, 
computed by {\footnotesize axionCAMB} \cite{2015PhRvD..91j3512H}).
In the lower panels, we show the evolution of $\epsilon$
for the same choices of $\dot{R}_{\mathrm{ini}}$ or $N$
as in the upper panels. 
\cref{Fg.fordotR} and \cref{Fg.forN} together show that
for higher values of $\dot{R}_{\mathrm{ini}}$ or $N$,
the initial amplitude of $\epsilon$ is higher,
leading to a more significant deviation of the complex SFDM overdensity
from the axion overdensity in their early-stage evolutions.
On the other hand, for smaller $\dot{R}_{\mathrm{ini}}$ or $N$, 
the initial energy flux is also smaller,
and thus the overdensity of a complex scalar field is closer to that of a real field.
These behaviors are not surprising:
when $\dot{R}_{\mathrm{ini}}$ is large, the complex scalar field starts in kination, 
and when $N$ is large, the complex field approaches the spintessence limit
(large asymmetry between the particles and the antiparticles).
In these cases, the complex SFDM is dynamically distinct from a real axion.
Conversely, the less the values of $\dot{R}_{\mathrm{ini}}$ and $N$,
the more the complex field mimics a real field \cite{2002PhRvD..65h3514A}.

\begin{figure}[H]
  \centering
  \includegraphics[width=7cm]{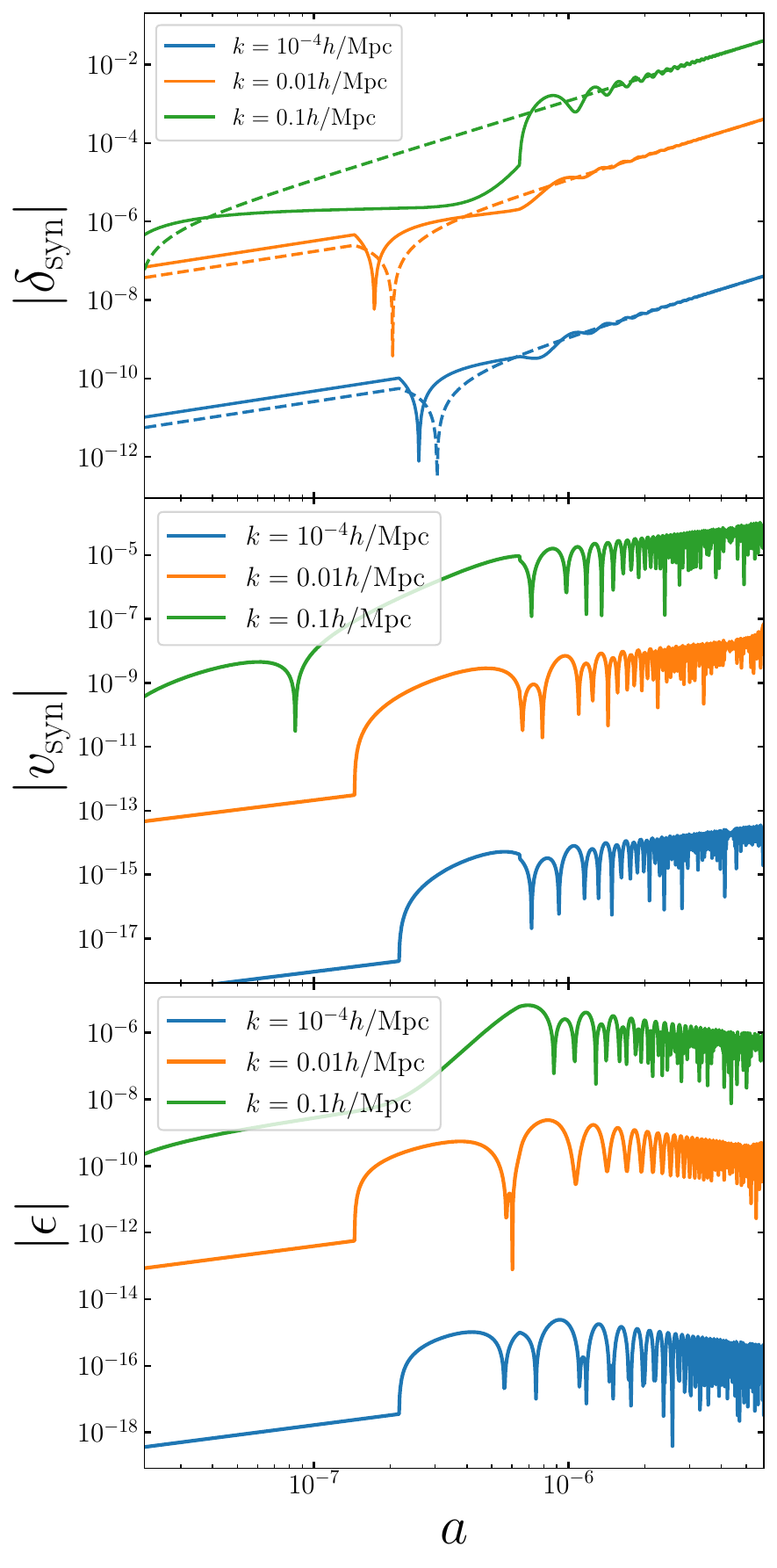}\\
  \caption{Early-stage evolutions of the SFDM fluid variables 
  at three wavenumbers.
  In the top panel, the solid curves denote the energy density contrasts
  and the dashed curves denote the charge density contrasts.
  Here we adopt $m = 10^{-22}\,\mathrm{eV}$, 
  $\dot{R}_{\mathrm{ini}} =10^{7}\,\mathrm{eV}^{2}$
  and $N = 10^{10}\,\mathrm{cm}^{-3}$}.\label{Fg.for_k}
\end{figure}





\begin{figure}[H]
  \centering
  \includegraphics[width=7.2cm]{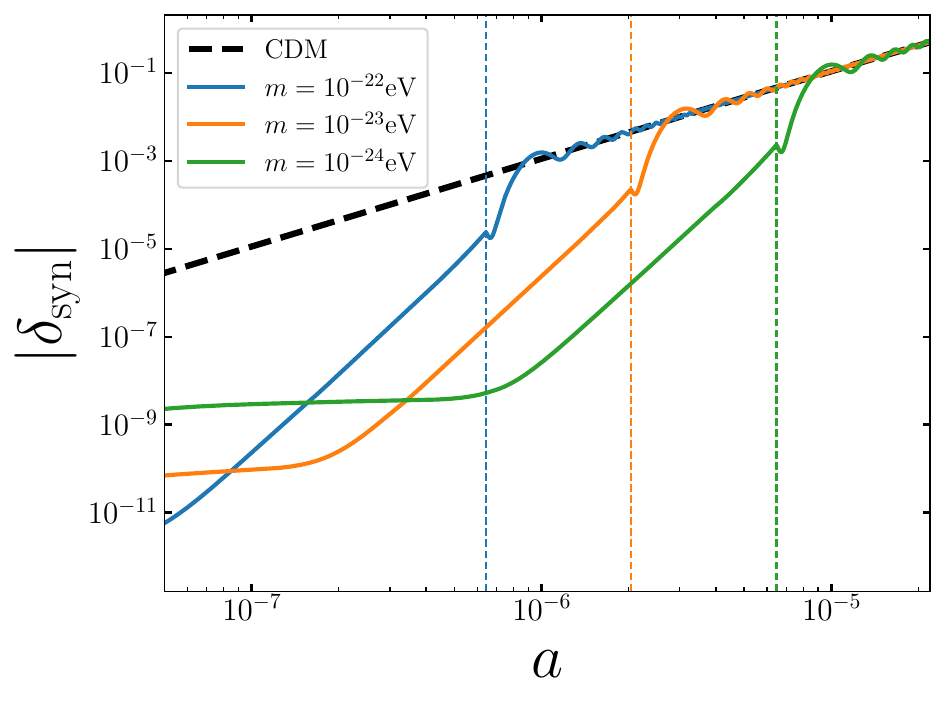}\\
  \caption{Early-stage evolution of the SFDM density contrast
  for various boson masses.
  Here we adopt 
  $\dot{R}_{\mathrm{ini}} =10^{4}\,\mathrm{eV}^{2}$
  and $N = 10^{10}\,\mathrm{cm}^{-3}$.
  The vertical dashed lines indicate the time of $a_\mathrm{osc}$ for each mass.}
  \label{Fg.delta_m}
\end{figure}

Finally, the scale dependence of the early-stage evolutions
of the SFDM fluid variables is illustrated in \cref{Fg.for_k},
where we show the evolutions of $\delta$, $\delta_n$, $v$ and $\epsilon$
at three different wavenumbers.
The effect of the boson mass
on the early-stage growth of the SFDM overdensity
is illustrated in \cref{Fg.delta_m}.
Similar to its effect on the background evolution, 
the boson mass impacts the evolution of the density contrast
mainly by inducing rapid oscillations at at $a> a_\mathrm{osc}$.

\section{Conclusion}\label{sec:conclusion}

In this paper, we established the theoretical ground
for future study of structure formation with the complex SFDM model,
in the framework of cosmological perturbation theory.
Based on the conserved Noether current
associated with the global U(1) symmetry,
we derived new hydrodynamic equations for complex SFDM
in the Eckart (particle) frame.
We showed that the complex SFDM can be effectively
described as an imperfect fluid with a dissipative energy flux.
Our numerical results for the background evolution recovered
the familiar behaviors of cosmological scalar fields.
The complex SFDM starts in a stiff/kination phase ($w=1$)
and then goes through a possible intermediate slow-roll phase ($w=-1$), 
before transitioning into the late-time oscillation regime.

For linear perturbations, 
we discovered a novel gauge condition related to the complex SFDM,
$c_s^2\equiv (\delta P/\delta\rho)|_s=1$, which we named the \emph{stiff} gauge.
Taking advantage of the stiff gauge,
we derived the constitutive equations for the complex SFDM, 
which are required to close the ODE system.
We modified the public Boltzmann code {\footnotesize CAMB}
to solve for the early-stage evolution of a $\Lambda$SFDM universe.
Using illustrative solutions,
we demonstrated the impact of the dissipative energy flux
on the early-stage growth of perturbations.
We discussed the differences between the perturbation evolution of the complex SFDM
and that of a real axion field, for various model parameters. 


In closing, it is worth noting that the formalism established in this paper
can be applied to any generic complex scalar field with a U(1) symmetry,
e.g., as in inflationary or dark energy models.
We look forward to investigating broader applications of the formalism
for future work.

\Acknowledgements{We thank the anonymous reviewers for their valuable comments.
This work was supported by the National Natural Science Foundation of China (Grant Nos. 12203012, 12494575) and Guangxi Natural Science Foundation (Grant No. 2023GXNSFBA026114).}

\InterestConflict{The authors declare that they have no conflict of interest.}



\bibliographystyle{unsrt}
\bibliography{refs}

\begin{appendix}




\renewcommand{\thesection}{Appendix}

\section{}

\subsection{\label{app:hydrovar}Hydrodynamic representation of complex SFDM}

For any continuous medium, we can define its hydrodynamic representation 
based on its stress-energy tensor, $T_{\mu\nu}$, 
as well as an arbitrary 4-velocity field, $u^\mu$, 
for a family of (fictitious) observers. 
The 4-velocity field induces the ``1+3'' splitting of spacetime with respect to
the rest frames of these observers,
via the projection tensor $h_{\mu\nu} \equiv g_{\mu\nu} - u_\mu u_\nu$. 
The decomposition of the stress-energy tensor 
results in the hydrodynamic variables in \cref{eq.2.1-6}, as follows:
\begin{equation}\label{eq.A-1}
  \begin{array}{rlrl}
    \rho & = ~T_{\mu\nu}\,u^{\mu}u^{\nu}, & P & = -\frac{1}{3}T_{\mu\nu}\,h^{\mu\nu}, \\
    q^\mu & = ~T_{\alpha\beta}\,h^{\mu\alpha}\,u^{\beta}, & \quad \pi^{\mu\nu} & = - T_{\alpha\beta}\,h^{\mu\alpha}\,h^{\nu\beta}-P\,h^{\mu\nu}.
  \end{array}
\end{equation}

Next, we decompose the complex scalar field
into a homogeneous part and a perturbed part,
and expand the stress-energy tensor defined in eq.~(\ref{eq.2.1-7}) to linear order.
The homogeneous part of $T_{\mu\nu}$ is given by
\begin{IEEEeqnarray}{rCl}
  \bar{T}_{00} & = & \dot{\bar{\phi}}^*\dot{\bar{\phi}}
  + m^2\,\bar{\phi}^*\bar{\phi}, \label{eq.BEMT1} \\
  \bar{T}_{0i} & = & \bar{T}_{i0} = 0, \qquad   \label{eq.BEMT2} \\
  \bar{T}_{ij} & = & a^2 \left(\dot{\bar{\phi}}^*\dot{\bar{\phi}} 
  - m^2\,\bar{\phi}^*\bar{\phi}\right) \delta_{ij},  \label{eq.BEMT3} 
\end{IEEEeqnarray}
and the perturbed part is given by
\begin{IEEEeqnarray}{rCl}
  \delta{T}^0_{~0} &=&    \dot{\bar{\phi}}\,\delta\dot{\phi}^* +\dot{\bar{\phi}}^*\delta\dot{\phi}  +m^2\left(\bar{\phi}\,\delta \phi^*+\bar{\phi}^*\,\delta \phi\right) -2\Psi\,|\dot{\bar{\phi}}|^2,\quad   \label{eq.PEMT1} \\
  \delta{T}^0_{~i} &=& \left(\dot{\bar{\phi}}\,\partial_i\delta \phi^* + \dot{\bar{\phi}}^*\,\partial_i\delta\phi\right),    \label{eq.PEMT2} \\
  \delta{T}^i_{~j} &=& -\Big[ \dot{\bar{\phi}}\,\delta\dot{\phi}^*
  +\dot{\bar{\phi}}^*\,\delta\dot{\phi} 
  -m^2\left(\bar{\phi}\,\delta \phi^*+\bar{\phi}^*\,\delta \phi\right)\nonumber\\
  & & - 2\Psi\,|\dot{\bar{\phi}}|^2\Big]\,\delta^i_{j}.  \label{eq.PEMT3} 
\end{IEEEeqnarray}
Inserting eqs.~(\ref{eq.BEMT1}-\ref{eq.BEMT3}) 
and the background component of the 4-velocity ($\bar{u}^0=1, \bar{u}^i=0$)
into eq.~(\ref{eq.A-1})
yields the homogeneous energy density and pressure of the complex scalar field,
as follows:
\begin{equation}\label{eq.A-2}
\begin{split}
   \bar{\rho} = & ~\dot{\bar{\phi}}^*\dot{\bar{\phi}}+ m^2\,\bar{\phi}^*\bar{\phi}, \\
   \bar{P}   =  & ~\dot{\bar{\phi}}^*\dot{\bar{\phi}} - m^2\,\bar{\phi}^*\bar{\phi}.
\end{split}
\end{equation}
It is also easy to see that both $\bar{q}^\mu$ and $\bar{\pi}^{\mu\nu}$
vanish at zeroth order.
Hence, the background complex SFDM is essentially a perfect fluid, 
as is any cosmic fluid.
For perturbations, combining eqs.~(\ref{eq.PEMT1}-\ref{eq.PEMT3})
and eqs.~(\ref{eq.2.1-10}-\ref{eq.2.1-12})
yields the following nonvanishing hydrodynamic variables:
\begin{IEEEeqnarray}{rCl}
     \delta \rho & = & ~\dot{\bar{\phi}}\,\delta\dot{\phi}^*
     +\dot{\bar{\phi}}^*\,\delta\dot{\phi} 
     -2\Psi\,|\dot{\bar{\phi}}|^2
     +m^2\left(\bar{\phi}\,\delta \phi^*+\bar{\phi}^*\,\delta \phi\right), \label{eq:delta_rho}\\
     \delta P & = & ~\dot{\bar{\phi}}\,\delta\dot{\phi}^*
     +\dot{\bar{\phi}}^*\,\delta\dot{\phi} - 2\Psi\,|\dot{\bar{\phi}}|^2
     -m^2\left(\bar{\phi}\,\delta \phi^*+\bar{\phi}^*\,\delta \phi\right), \qquad\label{eq:delta_P}\\
      Q_i & = & -(\bar{\rho}+\bar{P})\,(v_i-\partial_iB) 
      -\frac{1}{a}\left(\dot{\bar\phi}\,\partial_i\delta \phi^*
      +\dot{\bar\phi}^*\,\partial_i\delta \phi\right). \label{eq:energyflux}
\end{IEEEeqnarray}
Meanwhile, the anisotropic stress of the complex field
turns out to vanish at the linear order, $\Pi^i_{~j} = 0$,
since $\delta T^i_{~j} = - \delta P\,\delta^i_{~j}$ is diagonal
(cf. eq.~[\ref{eq.2.1-12}]).

Finally, we present the linear expansion of
the Noether current and the proper charge density
defined in eqs.~(\ref{eq.2.1-16})--(\ref{eq.2.1-18}).
The relationship between the charge density and the complex field
is expressed as follows:
\begin{IEEEeqnarray}{rCl}
  \bar{n} & = & \bar J^0 = 
  i \left(\bar{\phi}^*\,\dot{\bar{\phi}} - \bar{\phi} \,\dot{\bar{\phi}}^* \right), \label{eq:nbar}\\
  \delta n & = & - \bar{n}\,\Psi + 
  i\left(\bar{\phi}^* \delta \dot{\phi} - \bar{\phi}\,\delta \dot{\phi}^* 
  +\dot{\bar{\phi}}\,\delta \phi^* - \dot{\bar{\phi}}^*\,\delta\phi\right). \qquad \label{eq.A-4}
\end{IEEEeqnarray}
According to \cref{eq:fourvelocity},
the velocity of the charge flow can also be expressed in terms of the field,
\begin{equation}\label{eq.A-5}
    v_i = \partial_i B- \frac{i}{\bar n\,a} 
    \left(\bar{\phi}^*\,\partial_i \delta\phi 
    -\bar{\phi}\,\partial_i\delta \phi^*\right).
\end{equation}
Inserting the above equation into eq.~(\ref{eq:energyflux})
yields a nonvanishing energy flux, $Q_i$.
This is a direct result of our choice
of the 4-velocity associated with the Noether 4-current.

In general, the imperfect fluid nature (of complex SFDM)
is defined by the \emph{impossibility} of choosing a 4-velocity field
such that $q^\mu=\pi^{\mu\nu}=0$, as we mention in the main text.
By contrast, a real axion field with $\phi=\phi^*$ 
has a simple expression of the 4-velocity for the Landau-Lifshitz frame
(in which $q^\mu=0$ and $T^{\mu}_{~\nu}\,u^\nu=\rho\,u^\mu$):
$u_\mu=\partial_\mu\phi/\sqrt{\partial^\alpha\phi\,\partial_\alpha\phi}$
(up to a sign flip to make sure it is future-directed) \cite{1988CQGra...5..627M}.
It is easy to verify that in this Landau-Lifshitz frame,
the anisotropic stress $\pi_{\mu\nu}$ also vanishes,
and hence a real scalar field is a perfect fluid.
Note that the Noether 4-current defined in eq.~(\ref{eq.2.1-16}) becomes trivial
in the real field case, as expected.

\subsection{Derivation of the constitutive equations}\label{app:identities}

Here we present the derivation of the constitutive equations
as well as some useful identities.  
Rearranging \cref{eq.A-2} leads to the following relations:
\begin{equation}\label{eq.B-1}
  \begin{split}
     \bar{\rho} + \bar{P} = & ~2|\dot{\bar{\phi}}|^2,   \\
     \bar{\rho} - \bar{P} = & ~2 m^2\,|\bar{\phi}|^2.
  \end{split}
\end{equation}
These relations further yields
\begin{equation*}
  \begin{split}
      &\bar{\rho}^2 - \bar{P}^2 = 4 m^2|\dot{\bar{\phi}}|^2 |\bar{\phi}|^2, \\
      & \dot{\bar{\rho}} - \dot{\bar{P}} = 2 m^2 \frac{\ud}{\ud t} |\bar{\phi}|^2 
      = 2m^2 \left(\dot{\bar{\phi}}^*\,\bar{\phi} 
      + \bar{\phi}^*\,\dot{\bar{\phi}}\right).
  \end{split}
\end{equation*}
Combining the above equations with eq.~(\ref{eq:nbar}) yields
\begin{equation}\label{eq.B-2}
  \bar{\rho}^2 -\bar{P}^2 
  = m^2\bar{n}^2+\frac{(\dot{\bar{\rho}} - \dot{\bar{P}})^2}{4m^2}.
\end{equation}

In addition, we need the relations between 
the hydrodynamic variables for perturbations and the complex scalar field.
The expressions for the 3-velocity and the energy flux of complex SFDM are as follows:
\begin{IEEEeqnarray}{rl}
    & \bar{n}\,(v-B) = -\frac{i}{a} \left( \bar{\phi}^*\delta \phi -\bar{\phi} \delta\phi^* \right), \label{eq.B-3}\\
    & q = -(\bar{\rho}+\bar{P})(v-B) 
    -\frac{1}{a}(\dot{\bar{\phi}}\,\delta\phi^* + \dot{\bar{\phi}}^*\,\delta\phi). \label{eq.B-4}
\end{IEEEeqnarray}

Finally, to derive the constitutive equations in the stiff gauge, 
$\bar{\phi}\,\delta \phi^* + \bar{\phi}^*\,\delta \phi=0$,
we apply the following trick.
We define a real-valued function $\mathcal{K}$ that satisfies
\begin{equation}\label{eq.3.1-4}
  \frac{\delta \phi}{\bar{\phi}} = - \frac{\delta \phi^*}{\bar{\phi}^*}
  \equiv i\,\mathcal{K}.
\end{equation}
Then, eqs.~(\ref{eq.A-4}), (\ref{eq.B-3}), (\ref{eq.B-4}) 
can be reexpressed as follows:
\begin{IEEEeqnarray}{rl}
    & \delta n|_s + \bar{n}\,\Psi|_s= -2 |\bar{\phi}|^2\,\dot{\mathcal{K}}, \label{eq.3.1-5}\\
    & \bar{n}\,(v- B)|_s = \frac{2}{a}\,|\bar{\phi}|^2\,\mathcal{K}, \label{eq.3.1-6}\\
    & q|_s = -(\bar{\rho}+\bar{P}) (v - B)|_s + \frac{\bar{n}}{a}\,\mathcal{K}. \label{eq.3.1-7}
\end{IEEEeqnarray}

Combining the above equations, we can eliminate the auxiliary function $\mathcal{K}$
and obtain
\begin{equation}\label{eq.3.1-9}
  \frac{\delta n}{\bar{n}} + \Psi - 2aH(v -B) +a(\dot{v} - \dot{B}) 
  = \frac{\dot{\bar{\rho}}-\dot{\bar{P}}}{\bar{\rho}-\bar{P}}\,a(v- B)|_s.
\end{equation}
This relation is then used to derive eqs.~(\ref{eq.3.1-10}) and (\ref{eq.3.1-11}).

\subsection{Numerical method for the background solution}\label{app:numerical}

Here we describe how we integrate the homogeneous KG equation, 
eq.~(\ref{eq.2.1-4}), 
to obtain the background evolution of the complex scalar field. 
Integrating a complex scalar field is equivalent to solving 
for two real scalar fields, as $\bar\phi\equiv(\bar\phi_R+i\,\bar\phi_I)/\sqrt{2}$. 
The KG equations for two real scalar fields are two second-order ODEs. 
Thus, we need the array $\{ \bar{\phi}_{R,{\mathrm{ini}}}, 
\dot{\bar{\phi}}_{R,{\mathrm{ini}}},
\bar{\phi}_{I,{\mathrm{ini}}}, \dot{\bar{\phi}}_{I,{\mathrm{ini}}}\}$
as the initial conditions for integrating the KG equations, 
where the subscript `ini' denotes the initial moment at $a_{{\mathrm{ini}}}$.
However, these four parameters in the initial conditions are not independent.
We rewrite the complex scalar field in terms of its amplitude and phase, 
$\bar{\phi} = R\,e^{i\theta}$, and then recast \cref{eq.2.1-21,eq.A-2} into
\begin{IEEEeqnarray}{rCl}
    N & = & -2R^2\,\dot{\theta}\,a^3, \label{eq.C-1}\\
    \bar{\rho} & = & \dot{R}^2 + R^2\dot{\theta}^2 + m^2 R^2. \label{eq.C-2}
\end{IEEEeqnarray}
Thus, the initial conditions can be equivalently converted to
$\{R_{\mathrm{ini}}, \dot{R}_{\mathrm{ini}}, \theta_{\mathrm{ini}}, \dot{\theta}_{\mathrm{ini}}\}$.
Since the initial phase angle, $\theta_{\mathrm{ini}}$, is not observable,
we can set $\theta_{\mathrm{ini}} = 0$.
Meanwhile, the value of $\dot{\theta}_{\mathrm{ini}}$ is algebraically
related to $\dot{R}_{\mathrm{ini}}$ and the conserved charge density, $N$,
via eq.~(\ref{eq.C-1}).
Finally, the energy density of the complex SFDM today
yields a constraint concerning $R_{\mathrm{ini}}$ and $\dot{R}_{\mathrm{ini}}$.
As a result, only one of them is independent.
We choose $\dot{R}_{\mathrm{ini}}$ as the independent parameter
in the initial conditions, whose unit is $\mathrm{eV}^2$.
It is sometimes referred to as the initial radial velocity,
which determines the initial kinetic energy of the complex scalar field.
For large enough $\dot{R}_{\mathrm{ini}}$, 
the field starts in the kination phase ($w=1$).

In summary, the homogeneous evolution of the complex SFDM
is determined by the following four parameters:
$\{m$, $N$, $\Omega_{\phi}h^2$, $\dot{R}_{\mathrm{ini}}\}$,
where $\Omega_\phi\equiv\bar{\rho}/\rho_\mathrm{crit}$
is the density fraction of the complex SFDM today.
As mentioned above, once $\Omega_{\phi}h^2$ and $\dot{R}_{\mathrm{ini}}$ are fixed,
the value of $R_{\mathrm{ini}}$ will be fixed.
In this work, we implement a shooting algorithm
to determine its value, as described below.

When $H\gg 3m$, complex SFDM enters the oscillating phase (the scale factor 
at which it begins to enter the oscillating phase is named as $a_{\mathrm{osc}}$ 
at $H \approx 3m$), which is not applicable to numerical methods with a minimum step 
length greater than the oscillating time scale of $m^{-1}$. We have to choose a 
special moment to end the precise numerical algorithm, and at this moment we use 
the WKB approximation \cref{eq.4.1-3}, to solve the field results (the corresponding 
scale factor is named as $a_{\mathrm{wkb}}$,
and $a_{\mathrm{wkb}}/a_{\mathrm{osc}} > \mathcal{O}(10)$,
to ensures that the complex SFDM has strictly entered the oscillating phase). 
When the WKB approximation is used, 
the universe can be approximately described by the $\Lambda \mathrm{CDM}$ model. 
We can then calculate the energy density of the complex SFDM
at $a = a_{\mathrm{wkb}}$. The resulting energy density, 
$\bar{\rho}_{\mathrm{wkb}} = \Omega_{\phi}\,\rho_c\,a_{\mathrm{wkb}}^{-3}$,
allows us to determine the initial conditions by the shooting algorithm.
We construct the following objective function:
\begin{equation}\label{eq.C-3}
  L(R_\mathrm{ini}) \equiv
  \log\left[\frac{\bar{\rho}\,(R_\mathrm{ini},a_{\mathrm{wkb}})}{\bar{\rho}_{\mathrm{wkb}}}\right],
\end{equation}
where $\bar{\rho}\,(R_{\mathrm{ini}},a_\mathrm{wkb})$
is the SFDM energy density resulting from the initial condition
$R_{\mathrm{ini}}$ at $a = a_{\mathrm{wkb}}$.
While the choice of objective functions is not unique, 
the logarithm function helps the algorithm converge quickly in practice.
We use the bisection method to find the zero point of the objective function
in \cref{eq.C-3}.
In this way, we obtain all of our initial conditions.

\end{appendix}

\end{multicols}
\end{document}